\documentclass[aps,prd,superscriptaddress,12pt,showpacs,notitlepage]{revtex4}
	\usepackage{graphicx}
	
\usepackage{epsf}
\usepackage{epsfig}
\usepackage{amssymb}

\newcommand{\be}{\begin{equation}}
\newcommand{\ee}{\end{equation}}
\newcommand{\bea}{\begin{eqnarray}}
\newcommand{\eea}{\end{eqnarray}}

\newcommand{\bb}{\bibitem}
 
\newcommand{\eqn}{\begin{eqnarray}}
\newcommand{\eqnx}{\end{eqnarray}}

\begin{document}
\title{The  vector BPS Skyrme model}
\author{C. Adam}
\affiliation{Departamento de F\'isica de Part\'iculas, Universidad de Santiago de Compostela and Instituto Galego de F\'isica de Altas Enerxias (IGFAE) E-15782 Santiago de Compostela, Spain}
\author{C. Naya}
\affiliation{Departamento de F\'isica de Part\'iculas, Universidad de Santiago de Compostela and Instituto Galego de F\'isica de Altas Enerxias (IGFAE) E-15782 Santiago de Compostela, Spain}
\author{J. Sanchez-Guillen}
\affiliation{Departamento de F\'isica de Part\'iculas, Universidad de Santiago de Compostela and Instituto Galego de F\'isica de Altas Enerxias (IGFAE) E-15782 Santiago de Compostela, Spain}
\author{A. Wereszczynski}
\affiliation{Institute of Physics,  Jagiellonian University,
Reymonta 4, Krak\'{o}w, Poland}


\begin{abstract}
We analyze the vector meson formulation of the BPS Skyrme model in (3+1) dimensions, where the term of sixth power in first derivatives 
characteristic for the original, integrable BPS Skyrme model
(the topological or baryon current squared) is replaced by a coupling between the vector meson $\omega_\mu$ and the baryon current. We find that the model remains integrable in the sense of generalized integrability and almost solvable (reducible to a set of two first order ODEs) for any value of the baryon charge. Further, we analyze the appearance of topological solitons for two one-parameter families of one vacuum potentials: the old Skyrme potentials and the so-called BPS potentials. Depending on the value of the parameters we find several qualitatively different possibilities. In the massless case we have a parameter region with no skyrmions, a unique compact skyrmion with a discontinuous first derivative at the boundary (equivalently, with a source term located at the boundary, which screens the topological charge), and Coulomb-like localized solitons. For the massive vector meson, besides the no-skyrmion region and a unique C-compact soliton, we find exponentially as well as power-like localized skyrmions. Further, we find (for a specific potential) BPS solutions, i.e., skyrmions saturating a Bogomolny bound (both for the massless and massive vector mesons) which are unstable for higher values of the baryon charge. The properties of the model are finally compared with its baby version in (2+1) dimensions, and with the original BPS Skyrme model, contributing to a better understanding of the latter.   
\end{abstract}

\maketitle 
\section{Introduction}
Among the effective field theory approaches to strong interaction physics at low energies, the Skyrme model \cite{skyrme} plays a prominent role. The primary fields in the Skyrme model are the pions, whereas baryons, nucleons and nuclei are described by collective nonlinear excitations of the fundamental degrees of freedom of the theory, that is, topological solitons. The Skyrme model is very successful in the qualitative description of physical properties of nucleons and nuclei. First of all, the requirement of finite energy field configurations leads to an effective one-point compactification of the (three-dimensional)  base space with the resulting topology of the three-sphere $S^3$. Field configurations may, therefore, be interpreted as maps from  this base space $S^3$ to the field space (the group manifold SU(2)), which are characterized by an integer-valued topological degree or winding number. In the Skyrme model, this winding number is identified with baryon number, which is known to be conserved to a high precision. Further, a collective coordinate quantization of some light degrees of freedom (concretely, spin and isospin) about classical soliton solutions may be performed \cite{AdNaWi}, such that baryons with odd baryon number are always quantized as fermions with half odd-integer spin and isospin, as obviously must hold true. The resulting quantum states may be identified with the nucleons and with both fundamental and excited states of nuclei, where the comparison with the experimentally measured spectra of nuclei leads to rather satisfactory results in those cases where a detailed calculation has already been performed (see e.g. \cite{massive3}). On a more quantitative level, the Skyrme model, nevertheless, has some known drawbacks. First and foremost, higher soliton solutions correspond to rather strongly bound one-soliton bound states, which is in conflict with the small binding energies of physical nuclei. This problem is related to the fact that, although there exists a BPS bound already for the original Skyrme model, nontrivial soliton solutions cannot saturate this bound.  The question of how to improve the Skyrme model towards an (almost) BPS theory is, therefore, an important issue. There exist two main known possibilities to improve this situation. One may modify the Lagrangian without altering its field contents, or one may introduce additional fields. The original Skyrme Lagrangian consists of two terms (the subindices refer to powers of first derivatives, and $U$ is a SU(2) matrix), 
\begin{equation}
{\mathcal L}={\mathcal L}_2 + \mathcal{L}_4 ,
\end{equation}
the so-called nonlinear sigma model term
\begin{equation}
\mathcal{L}_2=-\frac{f_{\pi}^2}{4} \; \mbox{Tr} \; (U^{\dagger} \partial_{\mu} U \; 
U^{\dagger}
\partial^{\mu} U  )
\end{equation}
and the quartic Skyrme term, 
\begin{equation}
\mathcal{L}_4= - \frac{1}{32 e^2}\; \mbox{Tr} \; ([U^{\dagger} \partial_{\mu}
U,U^{\dagger} \partial_{\nu} U]^2 ),
\end{equation}
which is required to
circumvent the standard Derrick argument for the non-existence of
static solutions.
A first obvious generalization is the inclusion of a potential term
\begin{equation}
\mathcal{L}_0 = -\mu^2 V(U,U^{\dagger}),
\end{equation}
which is usually introduced to provide a mass term for the pions. Secondly, if a proper Hamiltonian formulation is required (i.e., no higher than second powers in time derivatives), then the only possible further generalization is provided by the following sextic term (the baryon number current squared)
\begin{equation}
\mathcal{L}_6=\lambda^2 \pi^4 {B}_\mu {B}^\mu   ,
\end{equation}
where ${B}_\mu$ is the topological (or baryon number) current
\begin{equation} \label{curr}
 B^\mu = \frac{1}{24\pi^2}   \mbox{Tr} \; (\epsilon^{\mu
\nu \rho \sigma} U^{\dagger} \partial_{\nu} U \; U^{\dagger}
\partial_{\rho} U \; U^{\dagger} \partial_{\sigma} U)  .
\end{equation}
We remark that from the point of view of the Derrick scaling argument the sextic term is as good as the quartic Skyrme term.
The generalized model consisting of all four terms has been studied and applied to the phenomenology of nucleons and nuclei, too, although the resulting numerical calculations are quite involved \cite{JJGBC} - \cite{FloPie}. 
Recently it has been found that the restricted model consisting only of the potential and sextic terms (the so-called BPS Skyrme model) has a BPS bound and exact soliton solution saturating this bound \cite{BPS-Sk1}, \cite{BPS-Sk2}. These classical solutions, therefore, correspond to nuclei without binding energies and realistic, small binding energies may be introduced both by quantum corrections and by small contributions of additional terms \cite{BoMa}. The BPS Skyrme model has further intriguing mathematical properties like, e.g., infinitely many symmetries and conservation laws as a consequence of its generalized integrability \cite{gen-int}. These symmetries contain the symmetries of an incompressible ideal liquid and allow, therefore, to reproduce some features of the liquid drop model of nuclear matter.

Another possibility to overcome the shortcomings of the Skyrme model consists in the inclusion of further (e.g., vector) fields. 
Firstly, one may couple the Skyrme fields to the electromagnetic field. The correct electromagnetic coupling of the Skyrme model was first derived in  \cite{wit1}, and soliton solutions of the resulting Maxwell--Skyrme system have been studied in \cite{pie-tra}, \cite{radu-tra} (the equivalent problem for the baby Skyrme model in one dimension lower has been studied in \cite{GPS}, \cite{schr1}, and, for the BPS baby Skyrme model, in \cite{gaugedBPSbaby}). Secondly, one may couple the Skyrme fields to vector mesons.
In a recent investigation, a Skyrme theory coupled to an infinite tower of vector mesons was derived from an instanton holonomy in one dimension higher, where the exact BPS property of the resulting Skyrme theory may be traced back to the self-duality of the instantons \cite{Sut1}. 
The infinite tower of Kaluza--Klein vector modes induces a flow to a conformal BPS theory which is a Minkowski space version of the Sakai--Sugimoto model (for further developments and applications to baryonic physics see, e.g., \cite{MaOh1}). So one main difference between the BPS Skyrme proposals of \cite{Sut1} and of \cite{BPS-Sk1} resides in the different symmetries which are associated to the BPS property (conformal symmetry in the former case, volume-preserving diffeomorphisms in the latter case). Also, it is not obvious how to include a potential term into the conformal setting of \cite{Sut1}, and this issue has not yet been completely resolved, to the best of our knowledge.  
Any truncation to a finite number of vector mesons of the theory of \cite{Sut1} leads to a theory which is no longer exactly BPS, but has rather small binding energies.  

Among all the vector meson couplings to the Skyrme model, there is one which results in a Lagrangian which is rather similar to the sextic term (the baryon current squared) above, namely the so-called omega meson $\omega_\mu$. Indeed, the omega meson couples to the topological current, ${\cal{L}} \sim \omega_\mu B^\mu$, and upon integrating out the omega meson, the sextic term is recovered in the limit of infinite vector meson mass.  
The importance of the omega meson is also related to the fact that it transfers the physical effects of the chiral anomaly to baryonic matter and that it prevents the solitons from shrinking due to the appearance of a short-range repulsion in nuclear interactions \cite{YeeYi}. Moreover, it blocks the flow to a conformal theory.

We remark that
the omega meson may be integrated out also for finite or zero meson mass, resulting in a nonlocal effective self-interaction of the topological current with an integral kernel $K^{\mu\nu}$ of the Yukawa or Coulomb type,
$$
L_{\rm eff} = \int d^3 y B_\mu (x) K^{\mu\nu} (x-y) B_\nu (y) ,
$$
although we shall not pursue this approach in the present paper (i.e., we will always maintain the omega meson explicitly). 
The same nonlocal interaction induced by a Yukawa or Coulomb integral kernel has been investigated for the nonlinear Schroedinger equation in lower dimensions in \cite{HaZa}.  
The scaling behavior of the vector meson terms (the coupling term and the standard kinetic term) is, in fact, such that it stabilizes the solitons without the need for a Skyrme term \cite{vec1} - \cite{vec3}, and the resulting theory consisting of ${\cal L}_0$ and ${\cal L}_2$ coupled to the omega meson has been studied recently in \cite{Sutcliffe vector}, where it was found that the soliton solutions are quite similar to the solitons of the Skyrme model. 

It is the purpose of the present paper to study in detail the vector meson version of the BPS Skyrme model, which is obtained from the vector model described in the preceding paragraph by suppressing the sigma model term ${\cal{L}}_2$, and to compare its properties to the standard BPS Skyrme model. First of all, we shall find that the infinitely many symmetries of the BPS Skyrme model carry over almost unaltered to the vector BPS Skyrme model and, consequently, we will be able to perform most of the calculations analytically, like in the BPS Skyrme model case. Concerning soliton solutions, we shall find that the solitons of the vector BPS Skyrme model are rather different from the ones of the standard BPS Skyrme model, at variance with the results of \cite{Sutcliffe vector} for the full Skyrme model and its vector version. This different result is, in some sense, expected, at least for certain potentials. The reason is that for potentials which include a pion mass term, the suppression of the term ${\cal L}_2$ corresponds to the limit of infinite pion mass in the sense that linear fluctuations of the pion field are completely suppressed. 
On the other hand, physically the pions are the lightest effective particles. Hence, the inclusion of vector mesons with a finite mass and a standard kinetic term in the BPS Skyrme model in some sense reverses the typical mass hierarchy of low-energy QCD. It would, therefore, be surprising and cast some doubt on the viability of the BPS Skyrme model as an effective theory for strong interaction physics, if it lead to qualitatively similar solitons as in the vector model with its inverted mass hierarchy. We remark that the vector versions of the baby Skyrme model and its BPS restriction have been investigated in \cite{Sut-baby-vector} and in \cite{BPSbaby-vector}, respectively.
These results are extended and generalized to the 3+1 dimensional situation in the present paper (for the BPS case), and in \cite{Sutcliffe vector} (for the full Skyrme model case).

\section{The $\omega$-vector model}
The vector version of the BPS Skyrme model is given by the following Lagrange density
\begin{equation}
\mathcal{L}=-\mu^2 V(U, U^\dagger) - \frac{1}{4} (\partial_\mu \omega_\nu - \partial_\nu \omega_\mu)^2 +\frac{1}{2} M^2 \omega_\mu^2 +\lambda' \omega_\mu B^\mu
\end{equation}
where $B_\mu$ is the baryon current.

We use the standard parametrization of the $SU(2)$ chiral field
$$
U=e^{i \xi \vec{n} \cdot \vec{\sigma}}= \cos \xi +i\sin\xi \vec n \cdot \vec
\sigma, \;\;\; \vec n^2 =1, $$
where $\vec{\tau}$ are the Pauli matrices, $\xi$ is a real field and 
$\vec{n}$ is a unit three component vector field, which is further 
related to a complex field $u$ by means of the stereographic projection
$$ \vec{n}=\frac{1}{1+|u|^2} \left( u+\bar{u}, -i ( u-\bar{u}),
1- |u|^2 \right).
$$
Then, 
\begin{equation}
\mathcal{L}=-\mu^2 V(U, U^\dagger ) -  \frac{1}{4} (\partial_\mu \omega_\nu - \partial_\nu \omega_\mu)^2 +\frac{1}{2} M^2 \omega_\mu^2+\frac{i\lambda \sin^2 \xi}{(1+|u|^2)^2} \omega_\mu \epsilon^{\mu \nu \alpha \beta} \xi_\nu u_\alpha \bar{u}_\beta 
\end{equation}
where $\lambda$ is a new constant related to $\lambda '$. The potential term is assumed to depend only on ${\rm Tr}\, U$ i.e., on the scalar field $\xi$.  Concretely we are going to analyze in detail a family of potentials which provides a generalization of the usual Skyrme potential 
\begin{equation}
V=\left( \frac{1-{\rm Tr}\, U}{2} \right)^\alpha = \left( 1- \cos \xi \right)^\alpha .
\end{equation}
Soliton solutions of the Skyrme model with the old potential ($\alpha =1$) have been studied, e.g., in \cite{massive1}, \cite{massive2}, whereas generalized potentials were investigated, e.g.,  in \cite{gen massive1}, \cite{gen massive2}.
The pertinent field equations take the form
\begin{equation}
\partial_\mu  F^{\mu \nu} +M^2\omega^\nu +\frac{i\lambda \sin^2 \xi}{(1+|u|^2)^2} \epsilon^{\nu \alpha \beta \gamma} \xi_\alpha u_\beta \bar{u}_\gamma=0 \label{eq1}
\end{equation}
\begin{equation}
i\lambda \epsilon^{\alpha \beta \mu \gamma} \partial_\mu \omega_\alpha \xi_\beta \bar{u}_\gamma =0 \label{eq2}
\end{equation}
\begin{equation}
\frac{i\lambda \sin^2 \xi}{(1+|u|^2)^2} \epsilon^{ \alpha \mu \beta \gamma}  \partial_\mu \omega_\alpha  u_\beta \bar{u}_\gamma +\mu^2 V_\xi=0 . \label{eq3}
\end{equation}
We assume the natural static ansatz
\begin{equation}
\omega_0 \equiv \omega = \omega (r), \;\;\ \xi = \xi (r), \;\;\; u=v(\theta)e^{in\phi}
\end{equation}
and the other $\omega_i=0$. Then, their static versions are
\begin{equation}
\nabla^2_r\omega-M^2\omega=i\lambda \frac{\sin^2 \xi}{(1+|u|^2)^2} \nabla_r \xi (\nabla_\theta u \nabla_\phi \bar{u}-\nabla_\theta \bar{u} \nabla_\phi u) \label{eqstat1}
\end{equation}
\begin{equation}
\frac{i\lambda \sin^2 \xi}{(1+|u|^2)^2} \nabla_r \omega (\nabla_\theta u \nabla_\phi \bar{u}-\nabla_\theta \bar{u}  \nabla_\phi u ) + \mu^2V_\xi=0 . \label{eqstat2}
\end{equation}
Observe that one of the equations of motion i.e., (\ref{eq2}), is obeyed identically by the ansatz without any restrictions on the form of the ansatz functions. However, the ansatz is compatible with the remaining two equations only for a very restricted form of the complex field. Namely, we need that
\begin{equation}
\frac{\nabla_\theta u \nabla_\phi \bar{u}-\nabla_\theta \bar{u} \nabla_\phi u}{(1+|u|^2)^2} 
\end{equation}
is a function of $r$ only and does not depend on the angular variables. Then, the remaining equations (\ref{eqstat1}), (\ref{eqstat2}) become ODEs depending entirely on $r$. So, for our ansatz 
\begin{equation}
\frac{\nabla_\theta u \nabla_\phi \bar{u}-\nabla_\theta \bar{u} \nabla_\phi u}{(1+|u|^2)^2} =-\frac{2in}{r^2\sin \theta} \frac{vv_\theta}{(1+v^2)^2}
\end{equation}
where the function $v$ must obey the appropriate boundary conditions such that $u$ covers the target space two-sphere at least once. The well-known solution which covers the target space latitude exactly once is
\begin{equation}
v(\theta)=\tan \frac{\theta}{2} \;\;\; \Rightarrow \;\;\; \frac{\nabla_\theta u \nabla_\phi \bar{u}-\nabla_\theta \bar{u} \nabla_\phi u}{(1+|u|^2)^2} =-\frac{in}{2r^2} .
\end{equation}
Inserting this result into the static equations we get
\begin{equation}
\nabla^2_r\omega-M^2\omega=n\lambda \frac{\sin^2 \xi}{2r^2} \nabla_r \xi  \label{eq stat1}
\end{equation}
\begin{equation}
\frac{n\lambda \sin^2 \xi}{2r^2} \nabla_r \omega + \mu^2V_\xi=0 . \label{eq stat2}
\end{equation}
Further, we may use the last formula to eliminate the derivatives of the vector fields from the first equation. Then we arrive at a set of two first order ODEs of the gradient flow type 
\begin{equation}
 \frac{1}{r^2} \partial_r \left( r^4 \frac{2\mu^2}{n\lambda} \frac{V_\xi}{\sin^2 \xi} + \frac{n\lambda}{4} (\xi - \sin \xi \cos \xi) \right)  = - M^2 \omega  \label{flow1}
\end{equation}
\begin{equation}
 \omega_r=-r^2\frac{ 2\mu^2}{n\lambda} \frac{V_\xi}{\sin^2 \xi} . \label{flow2}
\end{equation}
The fact that the static equations of motion can be reduced to a set of two first order ODEs (solvability) may probably be related to the existence of infinitely many conserved charges (integrability).  This set of equations must be equipped with the proper boundary conditions which guarantee nontrivial topology,
\begin{equation}
\xi(r=0)=\pi, \;\;\;\; \xi(r=R_0)=0 ,
\end{equation}
where $R_0$ can be finite (compactons) or infinite for usual solitons. Further, the boundary conditions for the vector meson field are
\begin{equation}
\omega_r(r=0)=0, \;\;\;\; \omega (r=R_0)=0.
\end{equation}
Notice that the last formula leads to the following condition for the behavior of the profile function at the origin 
\begin{equation}
\lim_{r\rightarrow 0} \left(  r^2 \frac{ V_\xi}{\sin^2 \xi (r)} \right)=0 . \label{origin}
\end{equation}
\subsection{Massless case}
For the massless vector meson field one may integrate the first equation of motion completely 
\begin{equation}
r^4 \frac{2\mu^2}{n\lambda} \frac{V_\xi}{\sin^2 \xi} + \frac{n\lambda}{4} (\xi - \sin \xi \cos \xi)=C . \label{sol massless}
\end{equation}
The integration constant $C$ can be easily determined using the assumed boundary conditions and (\ref{origin}) 
\begin{equation}
C=\frac{\pi n \lambda}{4} .
\end{equation}
Thus, for the massless case we have obtained exact solutions (although usually it is not possible to write them in a closed form, i.e., as  $\xi =\xi (r)$) for any value of $n$. The questions whether these configurations correspond to a nontrivial topology and how the solutions are localized, are determined by the particular form of the potential. 
\subsubsection{No solutions - $\alpha \in [1, \frac{3}{2}) $}
For the most interesting case of the old potential the profile equation reads
\begin{equation}
r^4 \frac{2\mu^2}{n\lambda} \frac{1}{\sin \xi} + \frac{n\lambda}{4} (\xi - \sin \xi \cos \xi)=\frac{\pi n \lambda}{4} .
\end{equation}
However, the profile function defined by this equation cannot reach the vacuum $\xi=0$ for any value of $r \neq 0$. Indeed, the left hand side is singular at such a point while the right hand side is obviously finite. Thus, there are no topologically nontrivial configurations for the old potential. This can be generalized to all potentials with $\alpha <\frac{3}{2}$. Then, the left hand side of the profile equation still is singular at the vacuum $\xi=0$,
\begin{equation}
r^4 \frac{2\mu^2}{n\lambda} 2^{\alpha-2} \frac{1}{\cos \frac{\xi}{2}} \left( \sin \frac{\xi}{2} \right)^{2\alpha-3}  + \frac{n\lambda}{4} (\xi - \sin \xi \cos \xi)=\frac{\pi n \lambda}{4} .
\end{equation}
\subsubsection{Compacton - $\alpha = \frac{3}{2} $}
In this case we get
\begin{equation}
r^4 \frac{\sqrt{2}\mu^2}{n\lambda}  \frac{1}{\cos \frac{\xi}{2}} + \frac{n\lambda}{4} (\xi - \sin \xi \cos \xi)=\frac{\pi n \lambda}{4}
\end{equation}
which at the vacuum value of the profile function takes the form
\begin{equation}
R_0^4 \frac{\sqrt{2}\mu^2}{n\lambda} =\frac{\pi n \lambda}{4} .
\end{equation}
Obviously, we find a compact skyrmion for which the vacuum value must be reached at the finite radius
\begin{equation}
R_0 = \sqrt[4]{\frac{\pi n^2\lambda^2 }{4\sqrt{2} \mu^2 }} .
\end{equation}
As in the case of the vector BPS baby Skyrme model \cite{BPSbaby-vector}, any compacton solution is, in fact, a solution in the space {\bf C} of continuous functions, but not in the space {\bf C$^1$} of continuous functions with a continuous first derivative, because the derivative is discontinuous at the compacton boundary. Equivalently, any compacton is a solution of the field equations with a Dirac delta source located at the boundary of the compacton, which screens the topological charge generated in the interior of the skyrmion,
\begin{equation}
 \frac{1}{r^2} \partial_r \left( r^2 \omega_r + \frac{n\lambda}{4} (\xi - \sin \xi \cos \xi) \right)  =  -\frac{\pi n \lambda}{4} \delta (R-r) . \label{source}
\end{equation}
Hence, the total charge of this configuration is zero. 
\subsubsection{Coulomb-like localized solutions  - $\alpha > \frac{3}{2} $}
Now we find a power-like tail for the profile function of the skyrmion, where the power depends only on the potential but not on the topological charge
\begin{equation}
\xi = \left(  \frac{\pi n^2\lambda^2}{\alpha 2^{\alpha+1} \mu^2}
\right)^{\frac{1}{2\alpha-3}} \frac{1}{r^{\frac{4}{2\alpha-3}}}+...
\end{equation}
while the vector meson field is universally of the Coulomb type 
\begin{equation}
\omega_r= -\frac{\pi n \lambda}{4 r^2} +...
\end{equation}
The corresponding energy reads
\begin{eqnarray}
E = 4\pi \int_0^\infty r^2dr \left[ \mu^2V(\xi) - \frac{1}{2}\omega^2_r -\frac{\lambda n}{2r^2} \omega \sin^2 \xi \xi_r \right]  \\ =4\pi \int_0^\infty r^2dr\left[ \mu^2V(\xi) + \frac{1}{2}\omega^2_r \right] - 4\pi \int_0^\infty \frac{d}{dr} \left(r^2 \omega \omega_r \right) \\
=4\pi \int_0^\infty r^2dr\left[ \mu^2V(\xi) + \frac{1}{2}\omega^2_r \right]
\end{eqnarray}
where, using the equations of motion, we have combined the second and third term into a total derivative plus an additional term. Since, due to the boundary condition and the asymptotical form of the solution, the total derivative term is zero, we arrive at a positive definite expression for the total energy.  
Obviously, the second term is localized like a Coulomb electric field. On the other hand, the potential term converges in a way which depends on $\alpha$. Indeed, for $\xi \rightarrow 0$
$$\left. r^2  V(\xi)\right|_{on shell} = \left. r^2 (1-\cos \xi)^{\alpha} \right|_{on shell} \sim \left. r^2 \xi^{2\alpha}\right|_{on shell} \sim r^2 \left( \frac{1}{r} \right)^{\frac{8\alpha}{2\alpha -3}} =  \left( \frac{1}{r} \right)^{2 \frac{2\alpha+3 }{2\alpha -3}}.$$
It approaches the Coulomb like localization for $\alpha \rightarrow \infty$ while for any other $\alpha > 3/2$ we get a stronger convergence.  
\subsection{Massive case}
In the massive case, the field equation for the vector meson can be rewritten as
\begin{equation}
\partial_x \left[ (3x)^{4/3} \frac{2\mu^2 \alpha}{n\lambda} \frac{(1-\cos \xi)^{\alpha-1}}{\sin \xi} +\frac{n\lambda}{4} (\xi -\sin \xi \cos \xi) \right] = -M^2\omega \label{massive x}
\end{equation}
where, for simplicity, we introduced a new variable $x=r^3/3$. Then, knowing that 
\begin{equation}
\omega_x=-\frac{2\mu^2\alpha}{n\lambda}  \frac{(1-\cos \xi)^{\alpha-1}}{\sin \xi} \label{omega x}
\end{equation}
and acting with $\partial_x$ on (\ref{massive x}) we arrive at
\begin{equation}
\partial_x^2 \left[ (3x)^{4/3}  \frac{(1-\cos \xi)^{\alpha-1}}{\sin \xi} +\frac{n^2\lambda^2}{8\mu^2 \alpha} (\xi -\sin \xi \cos \xi) \right] =  M^2\frac{(1-\cos \xi)^{\alpha-1}}{\sin \xi} 
\end{equation}
In order to classify skyrmion solutions from the point of view of their asymptotic behavior it is enough to analyze this equation in the vicinity of the vacuum value $\xi=0$. Then we get
\begin{equation}
\partial_x^2 \left[ (3x)^{4/3}    \xi^{2\alpha-3} +\frac{n^2\lambda^2}{\mu^2 \alpha}  \frac{2^{\alpha-3}}{3} \xi^3 \right] =  M^2 \xi^{2\alpha-3} \label{xi0}
\end{equation}
Performing the series expansion we find three types of possible solutions:  a compacton, and exponentially as well as power-like localized skyrmions which are analyzed in the further part of this section.  
\\
Finally, let us notice that using the field equations the total energy may be rewritten in the following form
\begin{eqnarray}
&E&=4 \pi \int^\infty_0 r^2 dr \bigg[\mu^2(1-\cos \xi)^\alpha + \frac{1}{2} \omega_r^2 + \frac{1}{2} M^2 \omega^2 \bigg] \nonumber \\
&& = 4 \pi \int^\infty_0 r^2 dr \bigg[ \mu^2(2h)^\alpha + \frac{1}{2} \omega_r^2 + \frac{1}{2} M^2 \omega^2 \bigg] 
\end{eqnarray}
where 
\begin{equation}
h(r)=\frac{1}{2} (1-\cos \xi)
\end{equation}
\subsubsection{No solutions  - $\alpha < \frac{3}{2} $}
For $\xi$ close to the vacuum we get
\begin{equation}
\partial_x^2 \left[ (3x)^{4/3}    \xi^{2\alpha-3}  \right] =  M^2 \xi^{2\alpha-3} 
\end{equation}
where both the left and the right hand side diverge. Then, the asymptotic solution ($x \rightarrow \infty$) is
\begin{equation}
\xi^{2\alpha-3} \sim \frac{e^{+\sqrt[3]{3} M \sqrt[3]{x}}}{x^{4/3}}
\end{equation}
which means that $\xi$ is exponentially localized. However, it leads to an exponentially divergent vector meson (\ref{omega x}) and therefore to infinite energy. Hence, there is no skyrmion in this parameter range.  
\subsubsection{Compacton  - $\alpha = \frac{3}{2} $}
In this case one can find a compact solution
\begin{equation}
f(r)=\left\{ 
\begin{array}{cc}
\tilde{f}(x)& x \leq X \\
 & \\
 0 & x > X
\end{array} \right.
\end{equation}
\begin{equation}
\omega(x)=\left\{ 
\begin{array}{cc}
\tilde{\omega}(x) & x \leq X \\
 & \\
 0 & x > X
\end{array} \right.
\end{equation}
where $X$ is the (third power of the) radius of the compacton at which $\tilde{f}$ and  $\tilde{\omega}$ i.e., the solutions of (\ref{massive x}), (\ref{omega x}), are joined with the vacuum configuration.  However, $\omega_x$ has a jump at the boundary as $\tilde{\omega}_x(X)=-3\mu^2/n\lambda \sqrt{2}$. Hence, the derivative of it produces a delta like term located at the boundary. More precisely this boundary source term cancels the topological charge inside the compacton leading to a zero total charge configuration, exactly as in the massless case. 
\subsubsection{Exponentially localized skyrmions  - $\alpha \in ( \frac{3}{2}, 3] $}
Assuming $\alpha \leq 3$ one gets that $\xi^{2\alpha-3} \geq \xi^3$ for small $\xi$. Then, the first term in (\ref{xi0}) becomes the leading term under the additional condition (which must be verified at the end) that $\xi$ goes to 0 sufficiently fast. Hence, we find
\begin{equation}
\partial_x^2 \left[ (3x)^{4/3}    \xi^{2\alpha-3}  \right] =  M^2 \xi^{2\alpha-3} 
\end{equation}
which is a linear equation for $\xi^{2\alpha-3}$. It possesses an exponential solution  
\begin{equation}
\xi^{2\alpha-3} \sim \frac{e^{-\sqrt[3]{3} M \sqrt[3]{x}}}{x^{4/3}}
\end{equation}
Further, also the meson field and the energy density are exponentially localized. 
\begin{figure}[h]
 \begin{center}
  {\includegraphics[width=0.58\textwidth]{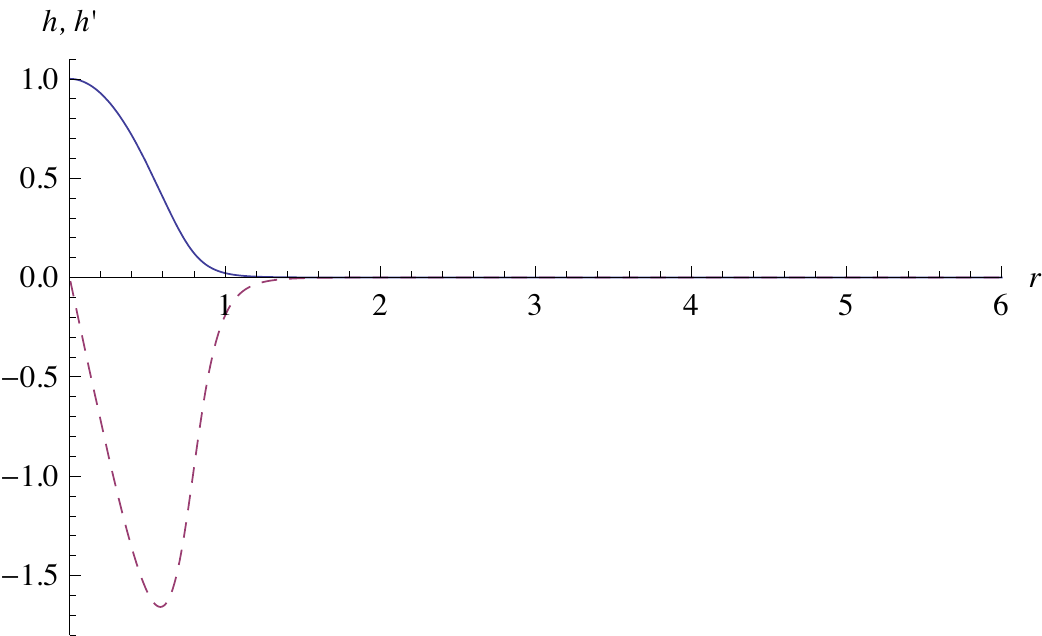}} 
 {\includegraphics[width=0.58\textwidth]{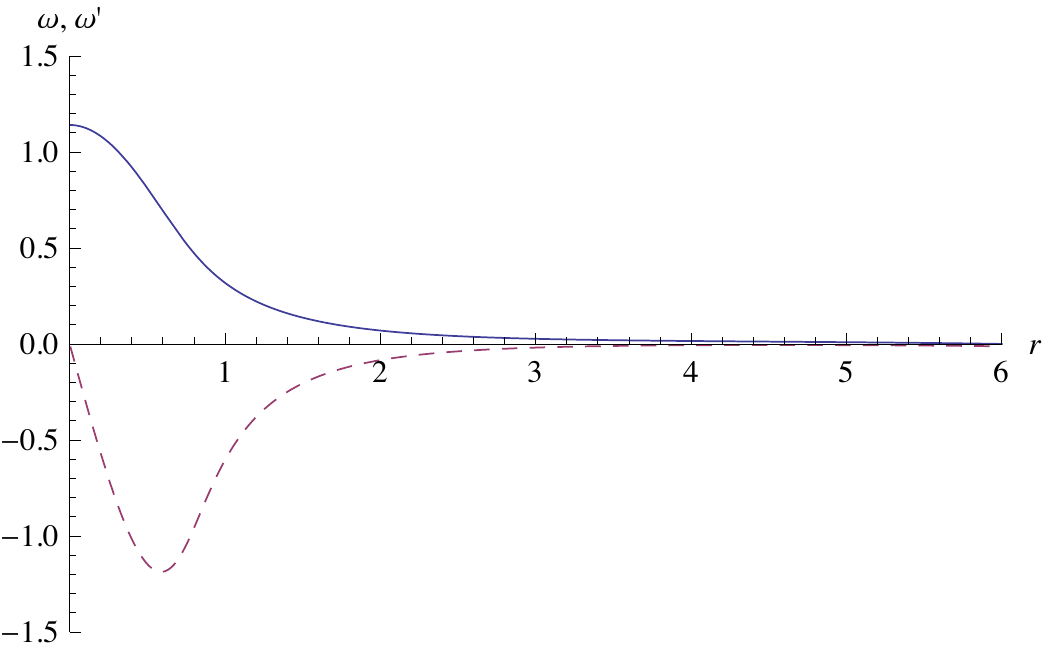}} 
 
 \vspace*{0.2cm}
 
 {\includegraphics[width=0.55\textwidth]{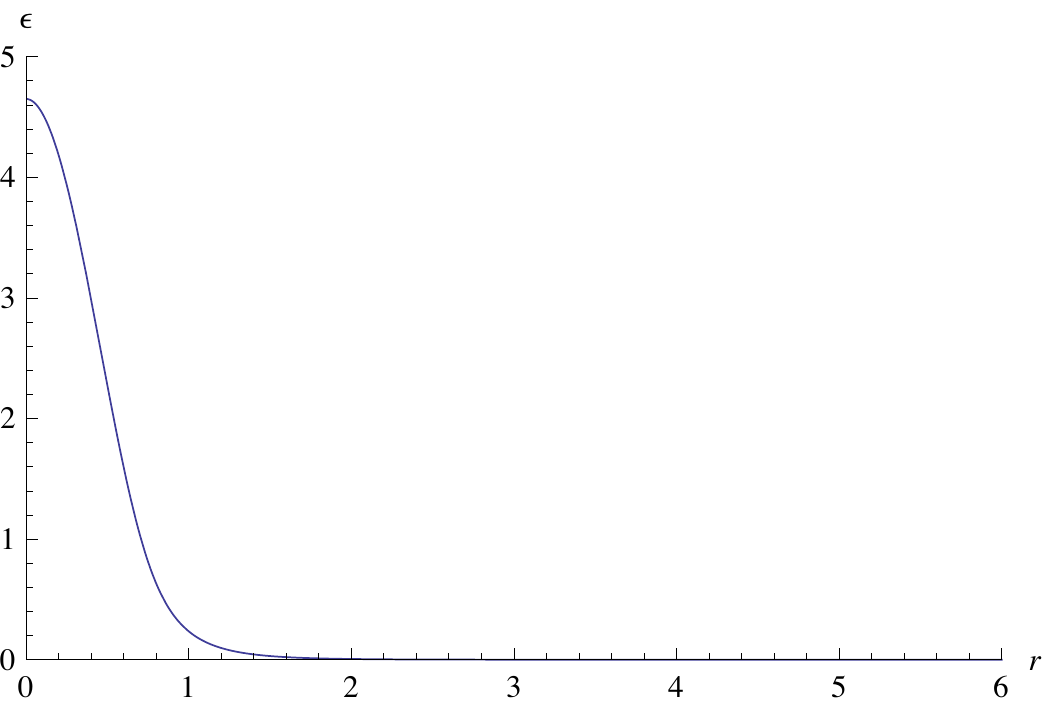}}
  \caption{Solutions with their derivatives and the energy density for the $\alpha =2$ massive case with the constants $\mu = M = \lambda = n = 1$.}
  \label{sola2}
 \end{center}
\end{figure}
\\
An example of a solution of this type has been calculated numerically for $\alpha=2$.  Then, performing the expansion around the center for the equations, we find (we use again the field variable $h$ instead of $\xi$)
\begin{equation}
h(r) \sim 1 + f_2 r^2 - \frac{M^2 \mu^2 f_2 - 5\mu^2 f_2^2 + n^2 \lambda^2 f_2^4}{5(\mu^2 + n^2 \lambda^2 f_2^2)} r^4 + ... 
\end{equation}
\begin{equation}
\omega(r) \sim v_0 + \frac{2 \mu^2 \sqrt{-f_2}}{n \lambda f_2} r^2 - \frac{\mu^2}{10 n \lambda \sqrt{-f_2}} \frac{M^2 \mu^2+6 n^2 \lambda^2 f_2^3}{\mu^2+n^2\lambda^2 f_2^2} r^4 +... 
\end{equation}
\noindent where $f_2$ and $v_0$ are the free parameters we can vary in order to find a numerical skyrmion solution with the correct asymptotic behavior for large $x$ (exponentially decreasing) via a shooting from the center.
For $\lambda=\mu=M=1$ and the simplest charge one skyrmion they take the following values 
\begin{equation}
f_2 = -1.8384364, \qquad \qquad v_0 = 1.14,
\end{equation}
\noindent and we find the solutions presented in Figure \ref{sola2}. The static energy corresponding to the energy density of Figure \ref{sola2} is
$
E=5.73645.
$
\subsubsection{Power-like localized skyrmions  - $\alpha > 3$}
For $\alpha >3$, the second term in  (\ref{xi0})  is dominating which leads to a different type of vacuum approach. Now we have
\begin{equation}
\frac{n^2\lambda^2}{\mu^2 M^2 \alpha}  \frac{2^{\alpha-3}}{3} \partial_x^2   \xi^3  = \xi^{2\alpha-3} 
\end{equation}
which gives a power-like localization
\begin{equation}
\xi \sim \left( \frac{1}{r} \right)^{\frac{1}{\alpha-3}} .
\end{equation} 
\begin{figure}[h]
 \begin{center}
{\includegraphics[width=0.58\textwidth]{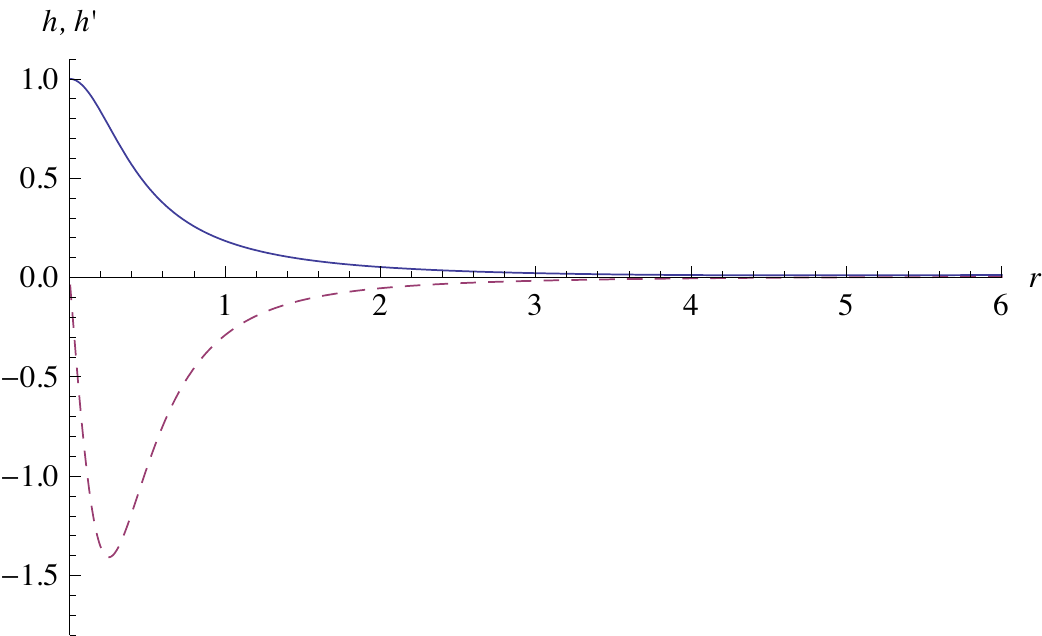}} \\
{\includegraphics[width=0.58\textwidth]{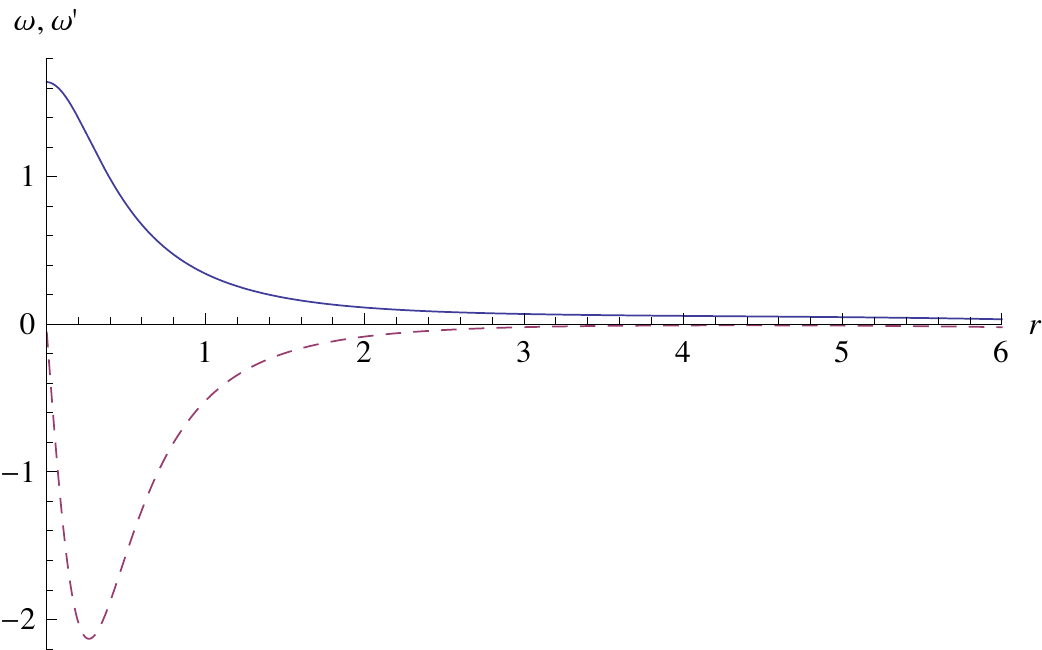}} 

\vspace*{0.2cm}

\includegraphics[width=0.55\textwidth]{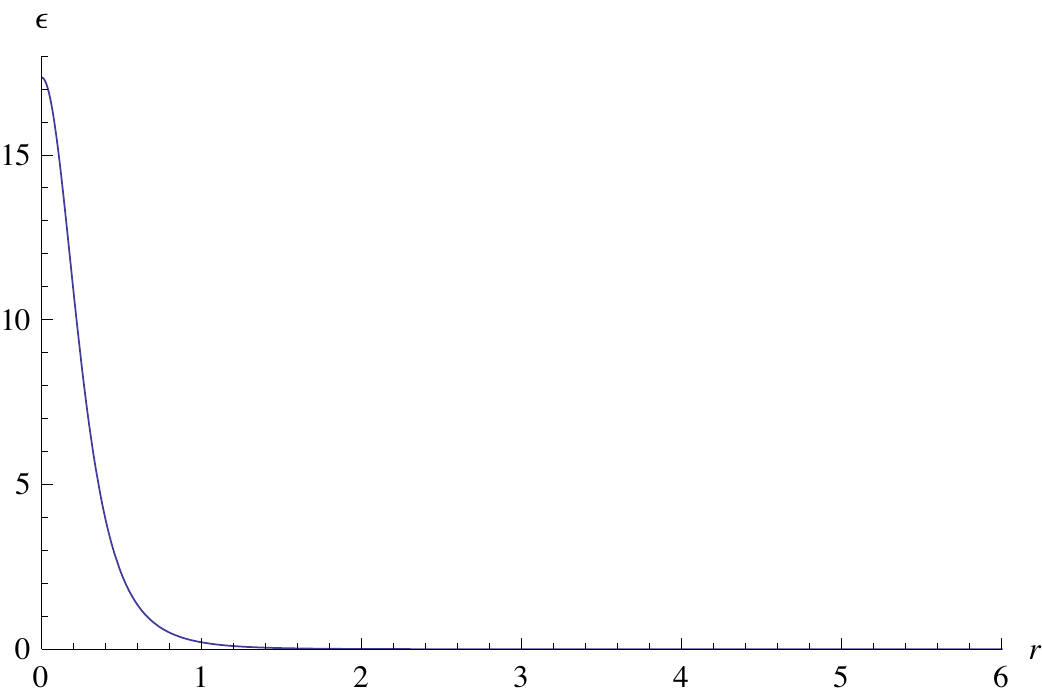}
  \caption{Solutions with derivatives and the energy density for the $\alpha =4$ massive case with the constants $\mu = M = \lambda = n = 1$.}
  \label{sola4}
 \end{center}
\end{figure}
\\
This behavior is reproduced by numerics. For example, for $\alpha=4$ after expanding around the center we find
\begin{equation}
h(r) \sim 1 + f_2 r^2 - \frac{8 M^2 \mu^2 - 200 \mu^2 f_2 + n^2 \lambda^2 f_2^3}{5(8 \mu^2 + n^2 \lambda^2 f_2^2)} f_2 r^4 +... ,
\end{equation}
\begin{equation}
\omega(r) \sim v_0 - \frac{16 \mu^2}{n \lambda \sqrt{-f_2}} r^2 - \frac{8}{5 \sqrt{-f_2}} \frac{8M^2 \mu^4 + 13 n^2 \lambda^2 \mu^2 f_2^3}{8n\lambda \mu^2 + n^3 \lambda^3 f_2^2} r^4 + ... ,
\end{equation}
The free parameters are now fixed as
\begin{equation}
f_2=-4.98624, \qquad \qquad v_0 = 1.64,
\end{equation}
\noindent which leads to the solution presented in Figure \ref{sola4} with the static energy
$
E=7.53696.
$
\subsection{BPS case}
Interestingly, there is a potential for which the energy density can be written as a total derivative. This potential possesses only one vacuum, where it looks similarly to the previously considered case $\alpha=4$. Globally, however, it has the following form
\begin{equation}
V_{BPS}=\frac{1}{4} \left( \xi - \cos \xi \sin \xi \right)^2 \label{potbps}
\end{equation}
It has the nice property that it allows to simplify one of the field equations
\begin{equation}
\frac{n\lambda}{2r^2} \sin^2 \xi \omega_r = - \mu^2 \left( \xi - \cos \xi \sin \xi \right) \sin^2 \xi \;\;\; \Rightarrow \;\;\;  \frac{n\lambda}{2r^2} \omega_r = - \mu^2 \left( \xi - \cos \xi \sin \xi \right) 
\end{equation}
or, using the variable $x=r^3/3$ introduced in Section II.B,
\begin{equation}
\frac{n\lambda}{2} \omega_x = - \mu^2 \left( \xi - \cos \xi \sin \xi \right) . \label{omegabps}
\end{equation}
Now, using the equation for the meson field, one can compute the total energy 
\begin{eqnarray}
E = 4\pi \int_0^\infty r^2dr \left[ \mu^2V_{BPS}(\xi) - \frac{1}{2}\omega^2_r  - \frac{1}{2}M^2\omega^2-\frac{\lambda n}{2r^2} \omega \sin^2 \xi \xi_r \right]  \\ =4\pi \int_0^\infty r^2dr\left[ \mu^2V_{BPS}(\xi) - \frac{\lambda n}{4r^2} \omega \sin^2 \xi \xi_r \right]\\
=4\pi \int_0^\infty dx \left[ \mu^2V_{BPS}(\xi) - \frac{\lambda n}{4} \omega \sin^2 \xi \xi_x \right] \\
=4\pi \int_0^\infty dx \left[ \mu^2 \left( \frac{n\lambda}{4 \mu^2} \right)^2  \omega_x^2 +  \frac{n^2 \lambda^2}{16 \mu^2} \omega \omega_{xx}  \right] 
\end{eqnarray}
where the last step follows from the derivative of  (\ref{omegabps}), i.e.,
\begin{equation}
\frac{n\lambda}{2} \omega_{xx} = -2\mu^2 \sin^2 \xi \xi_x .
\end{equation}
Then, combining the two terms into a total derivative we arrive at
\begin{equation}
E=\pi \frac{n^2 \lambda^2}{4\mu^2} \int_0^\infty dx \left[ \omega_x^2 +  \omega \omega_{xx}  \right] =\pi \frac{n^2 \lambda^2}{4\mu^2} \int_0^\infty dx \frac{d}{dx} ( \omega \omega_x)
 = -\pi \frac{n^2 \lambda^2}{4\mu^2} \omega(0) \omega_x(0)
\end{equation}
where the asymptotical vanishing of the meson field has been used. Further, the value of $\omega_x$ at the origin is fixed by equation (\ref{omegabps}) and the boundary condition for the profile function $\xi(0)=\pi$
\begin{equation}
\omega_x(0)=-\frac{2\pi \mu^2}{n\lambda} .
\end{equation}
Hence, finally 
\begin{equation}
E=\frac{\pi^2 n \lambda }{2} \omega(0) .
\end{equation}
Observe that this formula is valid for both the massless and the massive case, because the vector meson field is sufficiently localized also in its massless version. In the massless case, it is possible to find the value of the meson field at the origin analytically. The massless $\omega$ meson equation (\ref{sol massless}) has the solution
\begin{equation}
\xi - \cos\xi \sin \xi = \frac{\pi}{1+\frac{8\mu^2}{n^2\lambda^2}r^4} .
\end{equation}
Thus, 
\begin{equation}
\frac{n\lambda}{2} \omega_r = -  \frac{\pi \mu^2 r^2}{1+\frac{8\mu^2}{n^2\lambda^2}r^4}
\end{equation}
which may be integrated to the exact expression 
$$
\omega (r) = \frac{-2\pi \mu^2}{n\lambda} \frac{1}{4\sqrt{2}} \left( \frac{n\lambda }{2\sqrt{2}\mu} \right)^{3/2} \times 
$$
\begin{equation}
\left[ \ln \left( \frac{ar^2-\sqrt{2a} r+1}{ar^2+\sqrt{2a} r+1} \right) -2\arctan(1-\sqrt{2a} r) +2\arctan (1+\sqrt{2a}r) - 2\pi \right]
\end{equation}
where $a=2\sqrt{2} \mu / n\lambda$. Hence, its value at the origin is
\begin{equation}
\omega (0)= \frac{\pi^2}{4}  \left( \frac{n\lambda }{2\sqrt{2}\mu} \right)^{1/2} .
\end{equation}
Finally, the energy
\begin{equation}
E_{M=0}= \frac{\pi^4}{8\sqrt{2\sqrt{2}}} \lambda \sqrt{\frac{\lambda}{\mu}} \cdot n^{\frac{3}{2}}
\end{equation}
Obviously, a sum of $n$ charge one solutions is lighter than the charge $n$ soliton. Therefore, higher charge skyrmions are unstable although they saturate the BPS bound. Whether this is a consequence of the assumed spherical symmetry or a general property of this BPS model remains to be checked. 

\vspace*{0.2cm}

\begin{figure}[h]
\begin{center}
\includegraphics[width=0.6\textwidth]{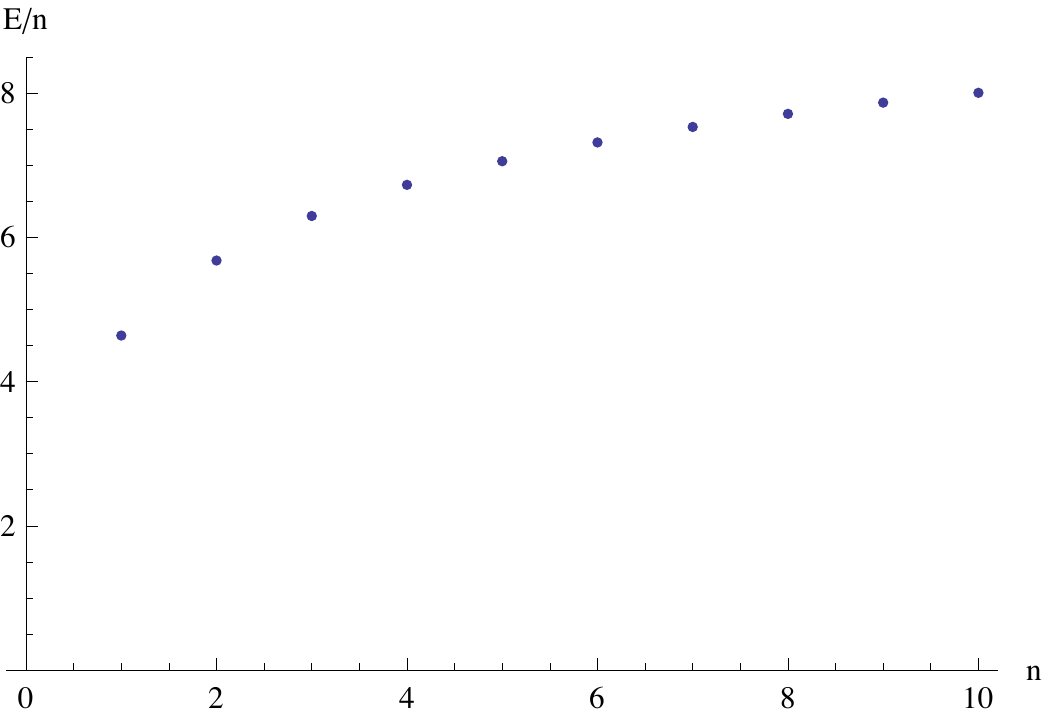}
\caption{Total energy over the topological charge as function of $n$.}
\label{energies}
\end{center}
\end{figure}

\begin{figure}[h]
 \begin{center}
\includegraphics[width=0.55\textwidth]{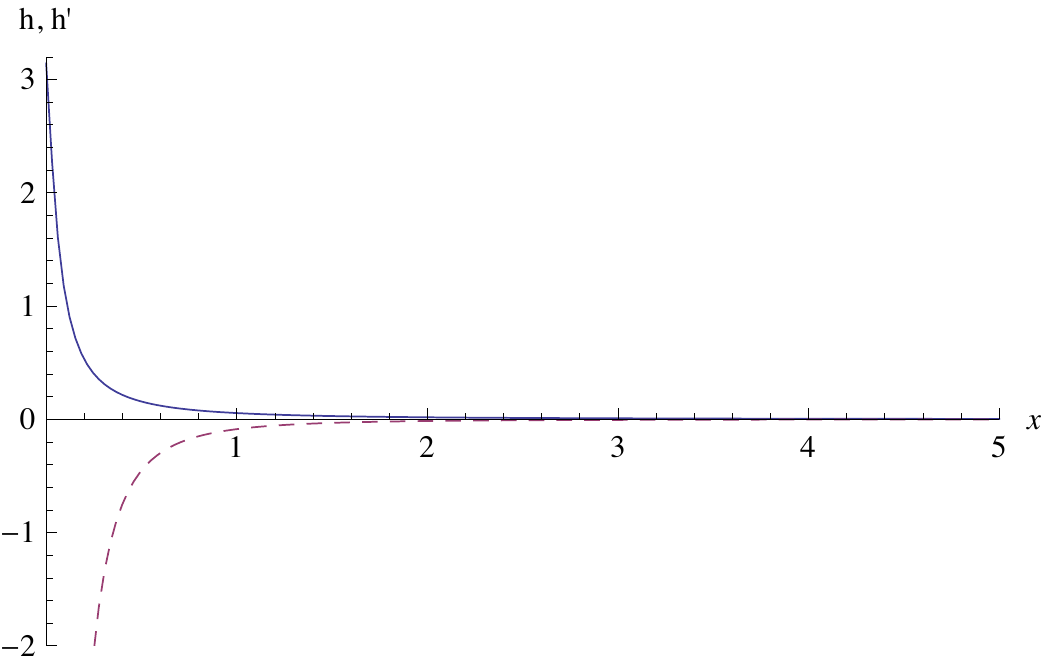} \\
\includegraphics[width=0.55\textwidth]{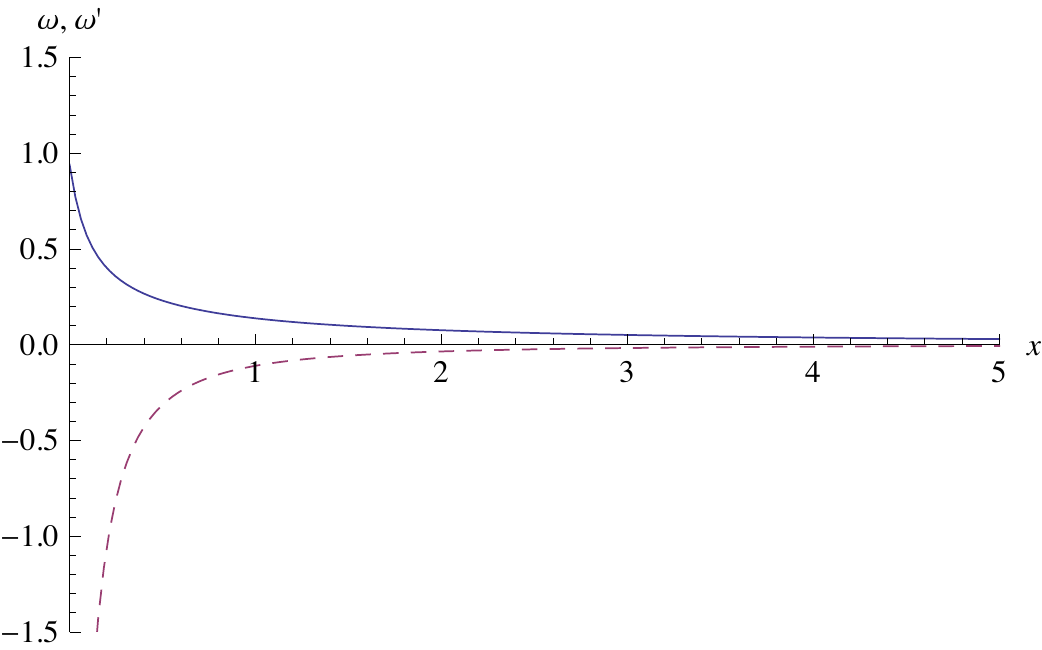} \\
\includegraphics[width=0.55\textwidth]{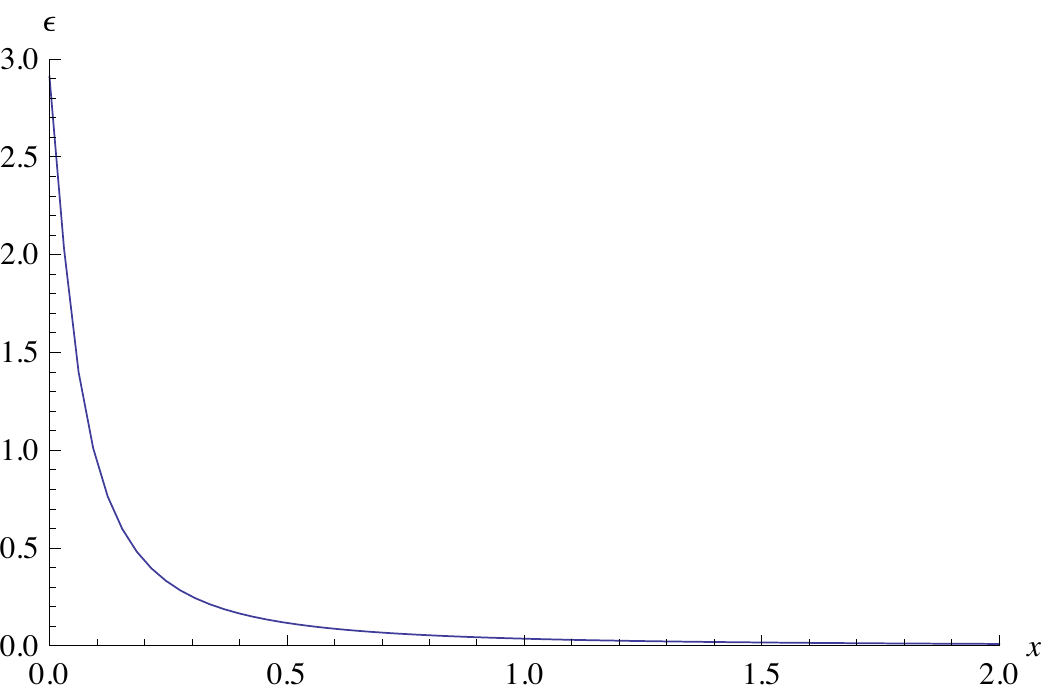}
  \caption{Solutions and energy density for the BPS potential with the constants $\mu = M = \lambda = 1$, and winding number $n=1$.}
  \label{BPS_Sol}
 \end{center}
\end{figure}

In order to (almost) analytically find the value of the meson field at the origin for the massive case we use its field equation 
\begin{equation}
 \frac{1}{r^2} \partial_r \left(\left(  r^4 \frac{2\mu^2}{n\lambda}  + \frac{n\lambda}{4} \right) (\xi - \sin \xi \cos \xi) \right)  = -M^2   \omega . \label{mass-omega}
\end{equation}
Then, we switch to the coordinate $x$ and differentiate the resulting equation by $\partial_x$. Using
 (\ref{omegabps}) we arrive at 
 \begin{equation}
 \partial_x^2 \left(\left(  (3x)^{4/3} \frac{2\mu^2}{n\lambda}  + \frac{n\lambda}{4} \right) (\xi - \sin \xi \cos \xi) \right)  = \frac{2M^2\mu^2}{n\lambda} (\xi - \sin \xi \cos \xi)   
\end{equation}
which is a second order but {\it linear} differential equation for $h=\xi - \sin \xi \cos \xi$
 \begin{equation}
 \partial_x^2 \left(\left(  (3x)^{4/3}  + \frac{n^2\lambda^2}{8 \mu^2} \right) h(x)  \right)  = M^2 h(x)
\end{equation}
The value of the vector meson field at the origin may be related to the derivative of $h$ by using formula (\ref{mass-omega})
\begin{equation}
\omega(0)=-\frac{n\lambda}{4M^2} h_x(0) .
\end{equation}
Unfortunately, the equation for $h$ can be solved only numerically, despite its linear character. We solved it by shooting from the center. We assume the standard values for the constants $\mu = M =\lambda = 1$. The range for the numerical solutions has  been chosen as $R=100$. Moreover, it has been checked that the BPS energy (computed for each skyrmion with $Q \in [1,10]$) gives exactly the same value like the one we get from the usual expression
\begin{equation}
E=4 \pi \int^\infty_0 dx \left[ \mu^2 \left( \frac{h}{2} \right)^2 + \frac{1}{2} (3x)^{4/3} \omega_x^2 + \frac{1}{2} M^2 \omega^2 \right] .
\end{equation}
It is clearly visible in Figure \ref{energies} that the higher charge skyrmions are unstable although they saturate the Bogomolny bound. Again, we observe a faster than linear growth of the energy with the topological charge. In fact, one can fit a curve $E=a \cdot |Q|^b$ and gets $a=5.13 \pm 0.07$ and $b=1.196 \pm 0.007$. Although we do not have any analytical argument, the power parameter $b$ is equal (within the error) to $\frac{6}{5}$. In any case, this dependence is slightly weaker than in the massless case. An example of solutions with $Q=1$ is presented in Figure \ref{BPS_Sol}. Taking the large $x$ limit, one can find that the solution possesses an exponential tail. Configurations for other topological charges look very similar. 

\vspace*{0.2cm}

Let us remark that for the family of potentials 
\begin{equation}
V=(V_{BPS})^\beta
\end{equation}
built on the BPS potential (\ref{potbps}), we observe the same pattern of solutions as in the previously analyzed case (the family of powers of the old potential). In the massless version, there are three possibilities: no skyrmions for $\beta < \frac{1}{2}$, a compacton with a screening source at the boundary for $\beta=\frac{1}{2}$, and Coulomb-type solutions for $\beta>\frac{1}{2}$. In the massive version we found four cases: no skyrmions for $\beta < \frac{1}{2}$, a compacton (which again requires a screening source) for $\beta=\frac{1}{2}$, and two infinitely extended types of solutions - exponentially ($\beta \in (\frac{1}{2}, 1]$) or power-like localized ($\beta >1$). In fact, this behavior repeats for any one-vacuum potential with the power-like approach to the vacuum at $\xi=0$ $$V \sim \xi^a$$
Here, $a=2\alpha$ for the old Skyrme potentials family or $a=6\beta$ for the BPS family. 
\vspace*{0.2cm}

All these results should be compared with the qualitative properties of skyrmions in the BPS Skyrme model for the potentials considered above. This is analyzed in the next section.
\section{Skyrmions in the BPS Skyrme model}
The BPS Skyrme model is \cite{BPS-Sk1}
 \begin{equation}
{\cal L}_{\rm BPS}=\lambda^2 \pi^4 B_\mu^2 - \mu^2 V(U,U^{\dagger}),
\end{equation}
which, using the field decomposition introduced previously, reads
\begin{equation}
{\cal L}_{\rm BPS}= -\frac{  \lambda^2 \sin^4 \xi}{(1+|u|^2)^4} \;\left(  
\epsilon^{\mu \nu \rho \sigma} \xi_{\nu} u_{\rho} \bar{u}_{\sigma} \right)^2
-\mu^2 V(\xi) .
\end{equation}
Assuming exactly the same Ansatz for the Skyrme field, we arrive at a differential equation for the profile function $\xi$ 
\begin{equation}
\frac{n^2\lambda^2 \sin^2 \xi }{2r^2} \partial_r \left(\frac{\sin^2 \xi \; 
\xi_r}{r^2} \right) - \mu^2 V_{\xi}=0 .
\end{equation}
This equation can be simplified by introducing the new variable $z$ (up to a numerical factor it is our previously defined $x$)
\begin{equation}
z=\frac{\sqrt{2}\mu r^3}{3 |n|\lambda} .
\end{equation} 
Then,
\begin{equation} \label{xi-eq}
\sin^2 \xi \; \partial_z \left(\sin^2 \xi \; \xi_z\right) -  V_{\xi}=0,
\end{equation}
and it may be integrated to 
\begin{equation} 
 \frac{1}{2} \sin^4 \xi \; \xi^2_z=V(\xi), \label{bps eq}
\end{equation}
\subsection{The old Skyrme potentials}
For the old Skyrme potentials it gives 
\begin{equation}
\pm \frac{1}{\sqrt{2}} \sin^2 \xi \xi_z=(1-\cos \xi)^{\alpha/2} .
\end{equation} 
In the generic situation, this equation is solved by a combination of hypergeometric functions which is not quite illuminating. However, some general observations can be easily made. 
\\
First of all, performing an expansion in the vicinity of the vacuum value one can identify for which potential compact skyrmions occur. Namely, if we assume that $\xi \approx 0$, then the BPS equation at the leading order reads
\begin{equation}
\pm \xi_z=2^{\frac{1-\alpha}{2}} \xi^{\alpha-2}
\end{equation} 
which possesses the obvious solution
\begin{equation}
\xi \sim \left\{
\begin{array}{cl}
(z-z_0)^{\frac{1}{3-\alpha}} & \alpha \in [1,3) \\
e^{-\frac{1}{2}z}  & \alpha =3 \\
\left( \frac{1}{z} \right)^{\frac{1}{\alpha-3}} & \alpha > 3 .
\end{array}
\right.
\end{equation} 
Thus, for $\alpha<3$ the resulting skyrmions are of the compact type. This means that the vacuum value is reached at a finite distance, the radius of the compacton. It is in agreement with the result for the old Skyrme potential (i.e., $\alpha =1$), for which compact configurations have been found previously. For $\alpha=3$, we have an usual exponentially localized solution, while for $\alpha>3$ a power-like approach to the vacuum is observed. One can find exact solutions which confirm this result. 
\begin{itemize}
\item[i) $\alpha=2$]
The pertinent solution reads
$$  \left\{
\begin{array}{lc}
 \xi + \sin \xi= -\sqrt{2} \left( z - \frac{\pi}{\sqrt{2}} \right)& z \in 
\left[0,\frac{\pi}{\sqrt{2}} \right] \\
\xi = 0 & z \geq \frac{\pi}{\sqrt{2}}.
\end{array} \right.
$$
As expected, we get a compact configuration. This solution has
a jump of derivatives at $z=\pi / \sqrt{2}$. 
In this case, the jump is finite, and, as in the case of the standard Skyrme potential, this jump is 
immaterial for physical quantities like the energy density or the topological
charge density. 
\item[ii) $\alpha=3$]
The solution is given by the following expression 
$$ \cos \frac{\xi}{2}+\ln \tan \frac{\xi}{4} = - \frac{z}{2}.$$
Notice that this profile function is not of compact nature but is non-zero 
for all $z$. The vacuum value is approached asymptotically at infinity.  
Moreover, this solution and its energy density are exponentially localized. 
\item[iii) $\alpha=4$]
Now, the solution fulfilling the assumed topological boundary condition is
$$\xi + 2 \cot \frac{\xi}{2} = \sqrt{2}\left(z+\frac{\pi}{\sqrt{2}} 
\right).$$
This solution is non-zero for all $z$ and $\xi'_z (z) \rightarrow 0$, 
if $z\rightarrow \infty$. It is localized like a polynomial in inverse powers
of $r$. 
\item[iv) $\alpha=6$]
For this value we derive again a solution which is 
non-zero for all $z$, 
$$ \xi (z)=2\; \mbox{arc cot} \sqrt[3]{3\sqrt{2}z}. $$ 
It easy to see that this configuration as well as the corresponding energy 
density are polynomially localized in inverse powers of $r$, as 
for the $\alpha=4$ solution. Indeed, 
$$ \varepsilon (z) = 16 \sqrt{2} \pi \mu \lambda n  \frac{1}{1+(3\sqrt{2} 
z )^{\frac{2}{3} } }.$$
\end{itemize}
Finally, we can compute the total energy of these solitonic configurations for any $\alpha$ value using the BPS property of the solutions. Then, in the $z$ variable
\begin{equation}
E=4\pi |n|\lambda \mu \int dz \xi_z \sin^2 \xi \sqrt{V} = 4\pi |n|\lambda \mu \int_0^{\pi} d\xi  \sin^2 \xi (1-\cos \xi)^{\frac{\alpha}{2}}
\end{equation} 
\begin{equation}
= 4\pi^{3/2}  2^{1+\frac{\alpha}{2}} \frac{\Gamma \left( \frac{3}{2}+\frac{\alpha}{2} \right)}{\Gamma \left(3+\frac{\alpha}{2}\right)} |n| \lambda \mu
\end{equation} 
The existence of compact, exponential and power-like skyrmions in the BPS Skyrme models is an expected phenomenon as it has its counterpart for a lower-dimensional version of the model - the so-called BPS baby Skyrme model.
\subsection{The family of BPS potentials}
Now we consider the family of potentials constructed from the previously introduced BPS potential
\begin{equation}
V=(V_{BPS})^\beta = \left( \frac{1}{4} (\xi -\cos \xi \sin \xi)^2 \right)^{\beta} .
\end{equation}
Locally near the vacuum, the family of BPS potentials looks exactly as the family of the old Skyrme potentials with the identification $\alpha=3\beta$. So, the approach to the vacuum is the same. On the other hand, the behavior at the origin where $\xi=\pi$ is different. However, this fact only weakly (quantitatively) influences the solutions. Hence, we may consider this family of potentials as an approximation of the more standard ones which allows for exact solutions for all values of the parameter $\beta$.
\\    
Now the BPS equation reads
\begin{equation}
\pm \frac{1}{\sqrt{2}} \sin^2 \xi \xi_z =  \left( \frac{1}{2} (\xi -\cos \xi \sin \xi) \right)^{\beta}
\end{equation}
It can be further simplified if we define a new target space variable
\begin{equation}
\eta =  \frac{1}{2} (\xi -\cos \xi \sin \xi)
\end{equation}
which interpolates between $\eta(0)=\frac{\pi}{2}$ and $0$. Observe that using this variable $V=\eta^{2\beta}$, and the profile equation
takes a very simple form
\begin{equation}
\pm \frac{1}{\sqrt{2}}  \eta_z = \eta^\beta .
\end{equation}
It can be easily solved leading again to three types of skyrmions. 
\\
For $\beta \in (0,1)$ we find compact skyrmions
\begin{equation}
\eta = \left\{
\begin{array}{lc}
 \left( \sqrt{2}(1-\beta) (z_0-z) \right)^{\frac{1}{1-\beta}} & z \in 
\left[0,z_0 \right] \\
0 & z \geq z_0.
\end{array} \right.
\end{equation}
For $\beta=1$ we have a unique exponentially localized solution
\begin{equation}
\eta=\frac{\pi}{2} e^{-\sqrt{2}z} .
\end{equation}
For $\beta >1$ we find power-like localized skyrmions
\begin{equation}
\eta= \left( \frac{1}{\sqrt{2}(\beta-1) (z+z_0)} \right)^{\frac{1}{\beta-1}} .
\end{equation}
Here $$z_0= \frac{1}{\sqrt{2} |\beta-1|} \left(\frac{\pi}{2} \right)^{1-\beta} . $$
Finally, the corresponding energy is
\begin{equation}
E=4\pi \frac{|n|\lambda\mu}{1+\beta} \left( \frac{\pi}{2} \right)^{1+\beta} .
\end{equation}
\section{Integrability and conservation laws}
As one might expect, the vector BPS Skyrme model is integrable in the sense of generalized integrability. In particular, there is an infinite family of conserved currents 
\begin{equation}
j_\mu^G=\frac{\delta G}{\delta \bar{u}} \bar{\mathcal{K}}_{\mu}-\frac{\delta G}{\delta u}\mathcal{K}_{\mu}
\end{equation}
where
\begin{equation}
\mathcal{K}^{\mu} = \epsilon^{\alpha \beta \nu \mu} \omega_\alpha \xi_\beta u_\nu, \;\;\;\; \bar{\mathcal{K}}^{\mu} = \epsilon^{\alpha \beta \mu \nu} \omega_\alpha \xi_\beta \bar{u}_\nu
\end{equation}
and $G=G(u,\bar{u}, \xi)$ is an arbitrary function of the target space coordinates.  Then,
\begin{equation}
\partial^{\mu} J_{\mu}^G= G_{\bar{u}\bar{u}} \bar{u}_{\mu} 
\bar{\mathcal{K}}^{\mu} + G_{\bar{u}u} u_{\mu} \bar{\mathcal{K}}^{\mu} + 
G_{\bar{u}} \partial_{\mu} \bar{\mathcal{K}}^{\mu} -G_{u \bar{u}}
\bar{u}_{\mu} 
\mathcal{K}^{\mu} -G_{uu} u_{\mu} \mathcal{K}^{\mu} - G_{u} \partial_{\mu} 
\mathcal{K}^{\mu}
\end{equation}
\begin{equation}
+ G_{\bar{u} \xi} \xi_{\mu} \bar{\mathcal{K}}^{\mu} - 
G_{u \xi} \xi_{\mu} \mathcal{K}^{\mu}=0
\end{equation}
where one has to use identities obeyed by $\mathcal{K}_{\mu}$
\begin{equation}
u_{\mu} \mathcal{K}^{\mu}=\xi_{\mu} \mathcal{K}^{\mu}=
0, \;\;\; \bar{u}_{\mu} \mathcal{K}^{\mu}=u_{\mu} \bar{\mathcal{K}}^{\mu}
\end{equation}
as well as the field equation
\begin{equation}
\partial_\mu \mathcal{K}^\mu=0 .
\end{equation}
Observe that this is exactly the same family of conserved quantities as for the original BPS Skyrme model. The interaction with the vector mesons does not spoil the generalized integrability property, which may perhaps be responsible for the solvability of our model. 
\\
The set of conserved currents is even bigger if one considers the massless version of the model. Now, we can construct the additional family of currents
 \begin{equation}
 j_\mu^H=H(u\bar{u}) F_{\mu \nu} ( \bar{u}u^\nu + u \bar{u}^\nu) = H(u\bar{u}) F_{\mu \nu} \partial^\nu ( \bar{u}u) 
\end{equation}
where $H$ depends now on the modulus of the complex scalar. Then,
\begin{equation}
\partial^\mu j_\mu^H = H F_{\mu \nu} \partial^\mu \partial^\nu ( \bar{u}u) + H' F_{\mu \nu}  \partial^\mu ( \bar{u}u)  \partial^\nu ( \bar{u}u) +H (\partial^\mu F_{\mu \nu}) \partial^\nu ( \bar{u}u) .
\end{equation}
The first two terms vanish due to the contraction of the antisymmetric tensor $F_{\mu \nu}$ with two symmetric ones. Using the field equation for the mesons we get
\begin{equation}
\partial^\mu j_\mu^H = -i\lambda H \epsilon^{\nu \alpha \beta} \frac{u_\alpha \bar{u}_\beta}{(1+|u|^2)^2}  ( \bar{u}u^\nu + u \bar{u}^\nu)=0.
\end{equation}
This family of conserved currents is identical to the one found in the baby version of the model \cite{BPSbaby-vector}. 
\section{Summary and conclusions}
In the present paper we investigated the vector BPS Skyrme model and its soliton solutions. In this model, the sextic term of the BPS Skyrme model is replaced by a coupling of the baryon current to a (massive or massless) vector field (the omega meson). First of all, we found that the resulting model is still integrable in the sense of generalized integrability \cite{gen-int} and possesses infinitely many symmetries and conservation laws, which are, in fact, identical to the ones of the original BPS Skyrme model. Further, we found that, for a spherically symmetric ansatz, the static field equations may be brought to a first order form of the evolutionary (or gradient flow) type, where this fact is probably related to the integrability properties of the model. For a massless vector meson, the resulting equations can even be integrated completely, although the solution can usually be given only in an implicit form. 

The soliton solutions of the vector BPS Skyrme model are quite different from the corresponding solutions of the original BPS Skyrme model. For the standard pion mass potential, for which the BPS Skyrme model has a compacton solution, the vector model has no soliton solution at all, because any local solution cannot be extended to a global one with the correct boundary conditions for a topological soliton (i.e., all formal solutions have infinite energy). As explained already in the introduction, this difference has to be expected, because the vector BPS Skyrme model with a pion mass term corresponds to a situation where the mass hierarchy is reversed with respect to the case of low energy QCD. 
This seems to imply that if one wants to include omega mesons in a physically reasonable way into the BPS Skyrme model (i.e., without altering the physical mass hierarchy), then the limiting case of infinite meson mass should be considered. But this brings us back exactly to the original BPS Skyrme model, up to a redefinition of its coupling constants. The BPS Skyrme model itself may, therefore, be interpreted as a vector meson Skyrme model in a certain limit, which might explain its BPS property from a slightly different perspective.

For other types of potentials, with a faster than quadratic approach to the vacuum, soliton solutions of the vector model may exist, but their behavior is quite different from the normal BPS Skyrme model case. Genuine compacton solutions, which are quite typical for the BPS Skyrme model, do not exist in the vector version. They only exist (for potentials with a cubic approach to the vacuum) for a modified field equation where an inhomogeneous delta function source term effectively screening the topological charge is introduced at the compacton boundary. In other words, the vector model has compacton solutions only in the space {\bf C} of continuous functions, but not in the space {\bf C$^1$} of continuous functions with a continuous first derivative. 
For potentials with an even faster (than cubic) approach to the vacuum, solitons with a power-like or exponential decay at spatial infinity can be found. Concretely, for a massless vector meson we find that the vector meson term in the energy density always localizes like a Coulomb term, whereas the (power-like) localization properties of the Skyrme field depend on the potential. In the massive vector meson case, depending on the vacuum approach of the potential, both terms may decay either exponentially or in a power-like fashion, as summarized in Fig. \ref{3-models} below.

\begin{figure}[h]
\begin{center}
\includegraphics[width=0.8 \textwidth]{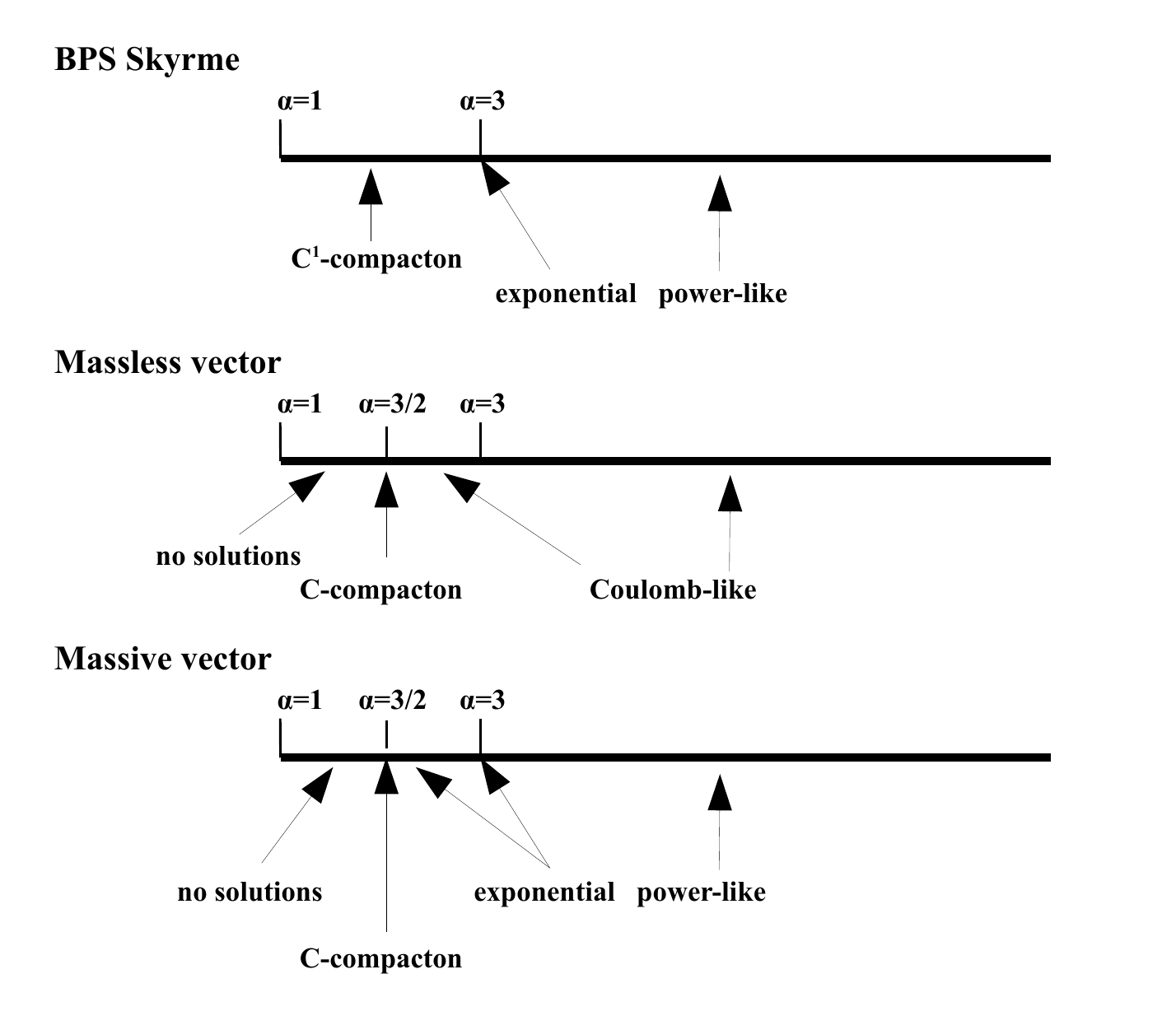}
\caption{Comparison of types of solutions in the BPS baby Skyrme mode (the old skyrme potentials) with its massless and massive vector counterparts. For the family of the BPS potentials the picture is exactly the same provided a substitution $\alpha \rightarrow 3\beta$ has been made.}
\label{3-models}
\end{center}
\end{figure}


Next, we considered the case of a specific potential (the BPS potential of Section II.C), where the energy density for spherically symmetric solutions may be written as a total derivative, such that these solutions saturate a BPS bound. It turns out that the energy of the BPS bound grows faster than linear in the topological charge and, therefore, higher charge solitons are unstable w.r.t. decay into smaller ones. At the moment it is not known whether this is just a property of the sector of spherically symmetric solitons, or whether this instability (and maybe even the BPS bound) continues to hold for the full model. Probably a full three-dimensional numerical simulation would be necessary to clarify this issue. In any case, this behavior is, again, completely different from the case of the BPS Skyrme model, where already the spherically symmetric solitons saturate a BPS bound linear in the topological charge. This linear behavior is, in fact, quite important for its applications to low-energy QCD and nuclear physics. For further recent results on nontrivial BPS bounds we refer, e.g., to \cite{Baz1}, \cite{gaugedBPSbaby}, \cite{BPSbaby-vector}.

If we compare our findings with the analogous results for the vector BPS baby Skyrme model in one dimension lower \cite{BPSbaby-vector}, then most of our results are quite similar. One minor difference is that in the baby case, the compacton solution with the effectively screened topological charge appears already for potentials with a quadratic approach to the vacuum (i.e., with a mass term). In both cases, genuine compactons do not exist for the vector models. This absence of compactons is probably related to the presence of a Coulombic (or Yukawa-like) term in the action of the vector models, which implies long-range interactions. Only if the topological charge (which plays a role analogous to the electric charge due to its coupling to the vector meson) is effectively zero because it is screened by a source term, these Coulomb or Yukawa long range interactions may be absent. It is interesting to observe that when the electromagnetic field is coupled to the baby Skyrme field in the standard, minimal way, then the resulting gauged BPS baby Skyrme model allows for compacton solutions \cite{gaugedBPSbaby} without problems. In this case, however, the electric field (and, therefore, the electric charge density) must be zero for finite energy configurations, and only magnetic fields are allowed. There is, therefore, no Coulomb term which would give rise to a long-range interaction.  

\vspace*{0.3cm}

{\centerline {\bf Acknowledgement}}

\vspace*{0.2cm}

The authors acknowledge financial support from the Ministry of Education, Culture and Sports, Spain (grant FPA2008-01177), the Xunta de Galicia (grant INCITE09.296.035PR and Conselleria de Educacion), the Spanish Consolider-Ingenio 2010 Programme CPAN (CSD2007-00042), and FEDER. CN thanks the Spanish Ministery of Education, Culture and Sports for financial support (grant FPU AP2010-5772). Further, AW was supported by polish NCN grant 2011/01/B/ST2/00464.

\end{document}